\documentstyle[twocolumn,floats,prb,epsfig,aps,amsmath,amssymb]{revtex}
\begin{document}
\draft

%\newfloat{figure}{ht}{aux} 
\preprint{IUCM96-019} 
\twocolumn[\hsize\textwidth\columnwidth\hsize\csname
@twocolumnfalse\endcsname

\title{Study of spin-Peierls transition in $\alpha ^{\prime
}$-NaV$_{2}$O$_{5}$ by infrared reflectivity}
\author{D.~Smirnov$^{a*}$, P.~Millet$^b$, J.~Leotin$^a$, 
D.~Poilblanc$^c$, J.~Riera$^d$, D.~Augier$^c$ and P.~Hansen$^d$}
\address{$^a$Laboratoire de Physique de la Mati\`{e}re Condens\'{e}e , 
SNCMP-INSA, Complexe de Rangueil, 31077 Toulouse, France}
\address{$^b$Centre d'Elaboration de Mat\'eriaux et d'Etudes  
Structurales CNRS, BP 4347,
29 rue Jeanne Marvig, 31055 Toulouse, France}
\address{$^c$Laboratoire de Physique Quantique \& UMR
CNRS 5626, Universit\'e P.~Sabatier,
31062 Toulouse, France\\
$^d$Instituto de Fisica Rosario, Consejo Nacional de Investigaciones 
Cientificas y Tecnicas y Departemento de Fisica,\\
Universidad Nacional de Rosario, Avemida Pellegrini 250, 2000-Rosario,
Argentina}
\date{\today}
\maketitle
\begin{abstract}

Polarized infrared reflectivity measurements have been 
performed on single crystals 
of the spin-Peierls compound $\alpha'$-NaV$_{2}$O$_{5}$~ in the 
temperature range 20--300 K.
Pronounced 
spectral features associated 
with the formation of the dimerized phase were detected both
in the ${\bf a}$- and ${\bf b}$-polarizations (perpendicular
and parallel to the spin-1/2 chains, respectively). The 
temperature dependence of 
a salient spectral line at 718~cm$^{-1}$ %(${\bf b}$-polarization)
sharply rising below the transition temperature $T_{\mathrm{SP}}$
obeys a $(1-T/T_{\mathrm{SP}})^{2\beta}$ law with 
T$_{SP}\simeq34.3$~K and $\beta\simeq0.25$. In addition, a continuum 
signal is 
observed in the whole temperature range in the ${\bf a}$-polarized
optical conductivity spectra. In order to interpret these results,
calculations of the
static dimerization and of the optical conductivity
based on a mean-field and a dynamical treatment of the lattice
respectively are proposed.
\end{abstract}
\pacs{78.30.Hv, 75.50.Ee, 63.20.-e, 64.70.Kb}
\vskip2pc]

The spin-Peierls (SP) transition has become a revived challenging effect in
quasi-one dimensional magnetism since the recent discovery of two inorganic
transition metal oxides compounds, first copper germanate 
CuGeO$_{3}$ (Ref.\onlinecite
{Hase}) and more recently sodium vanadate $\alpha ^{\prime }-$NaV$_{2}$O$_{5}$
(Ref.\onlinecite{Isobe}). These compounds 
contain linear spin-$1/2$ chains coupled to 
three-dimensional phonons. At 
low temperatures this coupling leads 
to the SP phase transition characterized by a
lattice dimerization together with the opening of a
spin gap.

The revival of the subject already investigated before in numbers of organic
salts\cite{Organic} is due to the availability of large size and high
quality crystals offering the possibility to carry out a number of relevant 
experiments~\cite{RevCuGeO3} including x-rays and neutron 
scattering~\cite{pouget,regnault,X-rays}.

Optical spectroscopy is known to be a sensitive tool to investigate 
properties of magnetic compounds\cite{Light} and
provides a detailed knowledge of phonons, magnons and their fluctuation
states. IR transmission and reflectance experiments
have been used to study the undistorted and SP phases in 
CuGeO$_{3}$ (Ref.\onlinecite{OptCuGeO3,li,Damasc}). The 
SP phase of NaV$_{2}$O$_{5}$ has been recently
studied by IR absorption in the frequency region 50--400\ cm$^ {-1}$. New
phonon lines have been found to accompany the SP 
transition\cite{Popova,Weiden}. Recently, the activation of a zone boundary
phonon in the SP phase of CuGeO$_{3}$ has been observed as a new single tiny
line in low temperature reflectivity spectra \cite{Damasc}. A distinctive
feature of NaV$_{2}$O$_{5}$ in comparison with CuGeO$_{3}$ is the negligible
magnetic interaction between the chains because of their isolation by V$^{5+}
$O$_{5}$ non magnetic chains. Frustration ratios $\alpha \approx 0.24\div
0.36$ (Ref.\onlinecite{riera,Castilla})
have been suggested for CuGeO$_{3}$ and $\alpha
\approx 0$ for NaV$_{2}$O$_{5}$ (Ref.\onlinecite{Isobe}). In addition, the
nearest-neighbor intrachain exchange constant J$\simeq$440 K, 
the transition temperature T$_{\mathrm{SP}}$= 35 K 
and the spin gap $\Delta_{\mathrm{spin}} 
$= 79 cm$^{-1}$ in NaV$_{2}$O$_{5}$ are typically 3-4 times larger than
in CuGeO$_{3}$ (Ref.\onlinecite{Weiden}). Also the  
dimerization $\delta$, estimated in the adiabatic approximation from 
the $T=0$ spin
gap, has the value $\delta\simeq 0.014$ (Ref.\onlinecite{riera})
and $\delta\simeq 0.048$
(Ref.\onlinecite{augier1}) for CuGeO$_{3}$ 
and NaV$_{2}$O$_{5}$, respectively. Thus, the
magneto-elastic coupling plays a larger role in NaV$_{2}$O$_{5}$
and enhanced spectral signatures of this coupling are expected in this
system.

In the following, we report on infrared reflectivity measurements of 
NaV$_{2}$O$_{5}$ single crystals over frequencies between 100 and 
10000~cm$^{-1}$ and
for temperatures between 20 and 300 K. The study focuses first on the room
temperature display of the excitation spectrum for the incident polarization
along the ${\bf a}$- and ${\bf b}$-axes (perpendicular
and parallel to the spin-1/2 chains, respectively). 
Then, prominent new lines giving
salient evidence of the dimerized phase below 34 K are reported. A detailed
study of the temperature dependence of the line at 718~cm$^{-1}$ is
presented. Finally, the outline of a calculation using Exact Diagonalization
(ED) and Quantum Monte-Carlo (QMC) algorithms based on a mean-field or a 
dynamical treatment of the lattice is given.

High-quality single crystals NaV$_{2}$O$_{5}$ with size up to 2.7$\times 
$5.1$
\times $0.3 mm$^{3}$ were prepared according to the procedure described in
Ref.\onlinecite{Techno}. 
%The samples were characterized by magnetic
%susceptibility measurements performed with a SQUID magnetometer
%at 3~K$\le T\le$350~K.
The optical measurements were done with a Bruker IFS 113V spectrometer at
nearly normal incidence of infrared radiation polarized along the ${\bf a}$-
and ${\bf b}$-axes and ${\bf k\parallel c}$. Room temperature transmission 
and reflection infrared spectra were measured in a spectral range
between 90 and 10000~cm$^{-1}$ with a resolution ranging from 1 cm$^{-1}$ in
the far-infrared region to 5 cm$^{-1}$ in the near-infrared region. The low
temperature far-infrared spectra were measured with a continuous He flow
cryostat down to 20 K with the absolute accuracy of temperature control
about 0.1~K. The spectral range was 90--700 cm$^{-1}$\ for ${\bf a}$%
-polarization and 90--1000 cm$^{-1}$ for ${\bf b}$-polarization with the
resolution 1~cm$^{-1}$.

We can distinguish three different regions in the measured infrared spectra
of NaV$_{2}$O$_{5}$ at 300 K. A rich phonon structure at frequencies below
1000 cm$^{-1}$ is followed by transmission windows 
at 1500--5150 cm$^{-1}$ for $%
{\bf a}$-polarization and at 1120--5320 cm$^{-1}$ for 
${\bf b}$-polarization. At higher frequencies, a wide maximum centered at
approximately 7100 cm$^{-1}$ was observed in the ${\bf a}$-polarized
reflectivity spectra. We have here restricted the discussion to the phonon
part of the spectra.

\begin{figure}[tbp]
\begin{center}
\epsfig{file=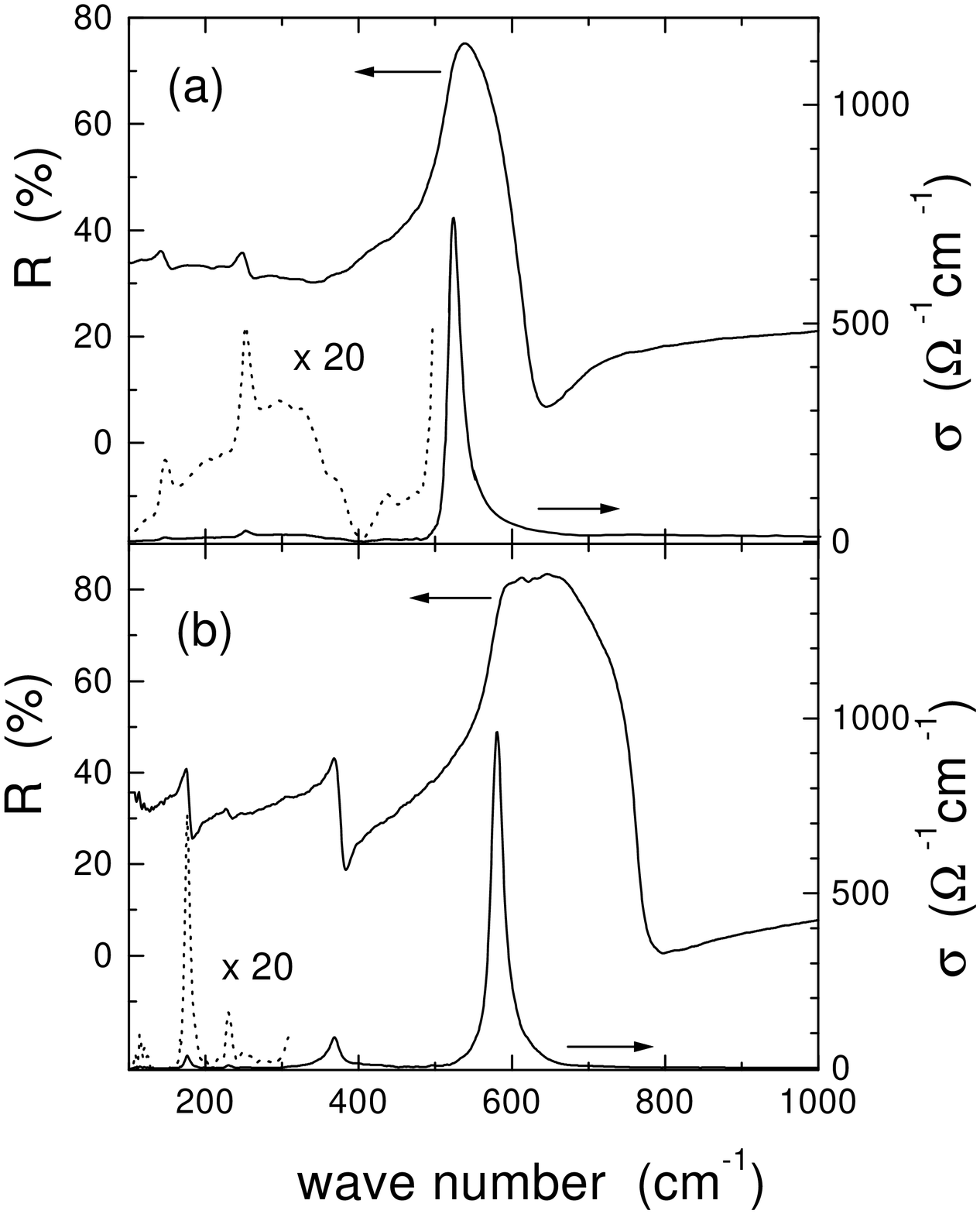,width=6cm}
\caption{Room temperature reflectivity spectra and optical
conductivity of NaV$_{2}$O$_{5}$ single crystal for ${\bf a}$-axis (a)
and ${\bf b}$-axis (b) polarizations.}
\label{refl}
\end{center}
\end{figure}

In Fig.~\ref{refl} the room temperature ${\bf a}$- and ${\bf b}$-axes
reflectivity spectra (R) and the frequency-dependent conductivity ($\sigma $)
derived via Kramers-Kronig analysis are shown in the region of lattice
vibration frequencies up to 1000 cm$^{-1}$. Three phonons are
detected for the ${\bf a}$-polarization ($\omega _{TO}\approx $ 524, 251,
145 cm$^{-1}$) and four for the ${\bf b}$-polarization ($\omega _{TO}\approx 
$ 581, 367, 229, 175 cm$^{-1}$). ${\bf b}$-axis polarized reflectivity
spectra have a typical form of phonon bands in an insulator which manifests
in dynamical conductivity as Lorentzian peaks. Obvious features in the 
${\bf a}$-polarized $\sigma $ spectrum are the occurrence of a broad excitation
band ending at 400~cm$^{-1}$ and an asymmetric line at 524 cm$^{-1}$ with an
underlying continuum extended above 1000 cm$^{-1}$ (Fig.~\ref{refl}a).
However, since the line shape depends strongly on the high-$\omega$
approximation of the data ($>10000$~cm$^{-1}$), reflectivity measurements 
in the visible and ultraviolet
regions are necessary for a definite analysis.

\begin{figure}[tbp]
\begin{center}
\epsfig{file=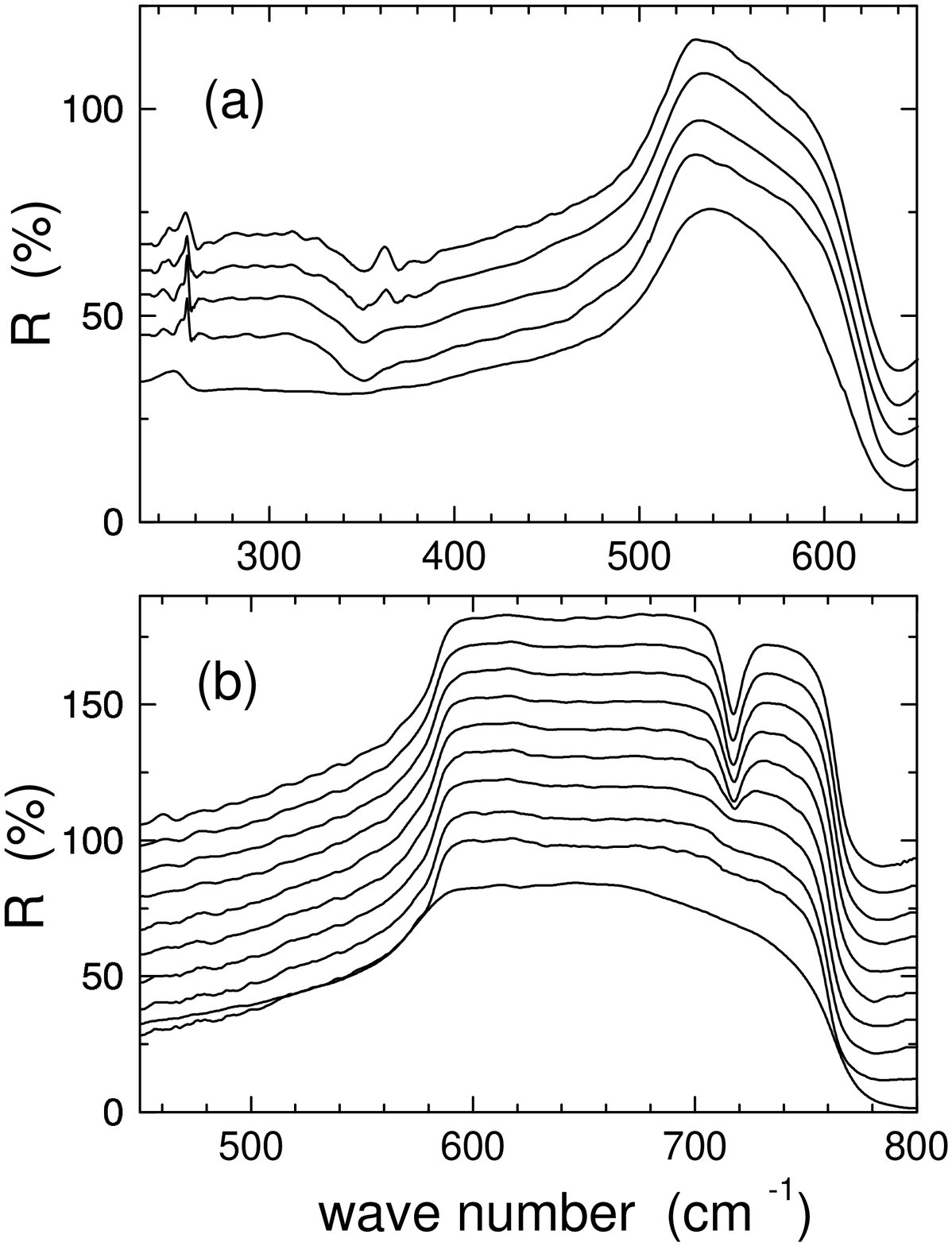,width=5.5cm}
\caption{Detailed temperature dependence in the reflectivity of $\protect%
\alpha ^{\prime }-$NaV$_{2}$O$_{5}$ single crystal. (a)-${\bf E\parallel a}$,
temperatures from up to down: 25, 29, 32, 37, 300K. (b)-${\bf E\parallel b}$,
temperatures from up to down: 21.8, 24, 27.1, 29.5, 32, 33.5, 34.3, 35.3,
38, 300K. Spectra are vertically shifted by 7\% (a) and 10\% (b).}
\label{temprefl}
\end{center}
\end{figure}

Below $T_{SP}$, a pronounced modification of
the reflectivity spectra is observed with sharply rising new features at
363, 200 cm$^{-1}$ (${\bf E\parallel a}$), 718~cm$^{-1}$ (${\bf E\parallel b}
$) and 138~cm$^{-1}$ (the polarization is unknown because of a poor
signal-to-noise ratio in the very far-infrared region). The ${\bf a}$-axis
polarized broad band at $\omega <400$ cm$^{-1}$ is enhanced at low
temperatures, but neither the position nor the intensity changes at 
$T=T_{SP}$ (Fig.~\ref{temprefl}).

\begin{figure}[tbp]
\begin{center}
\epsfig{file=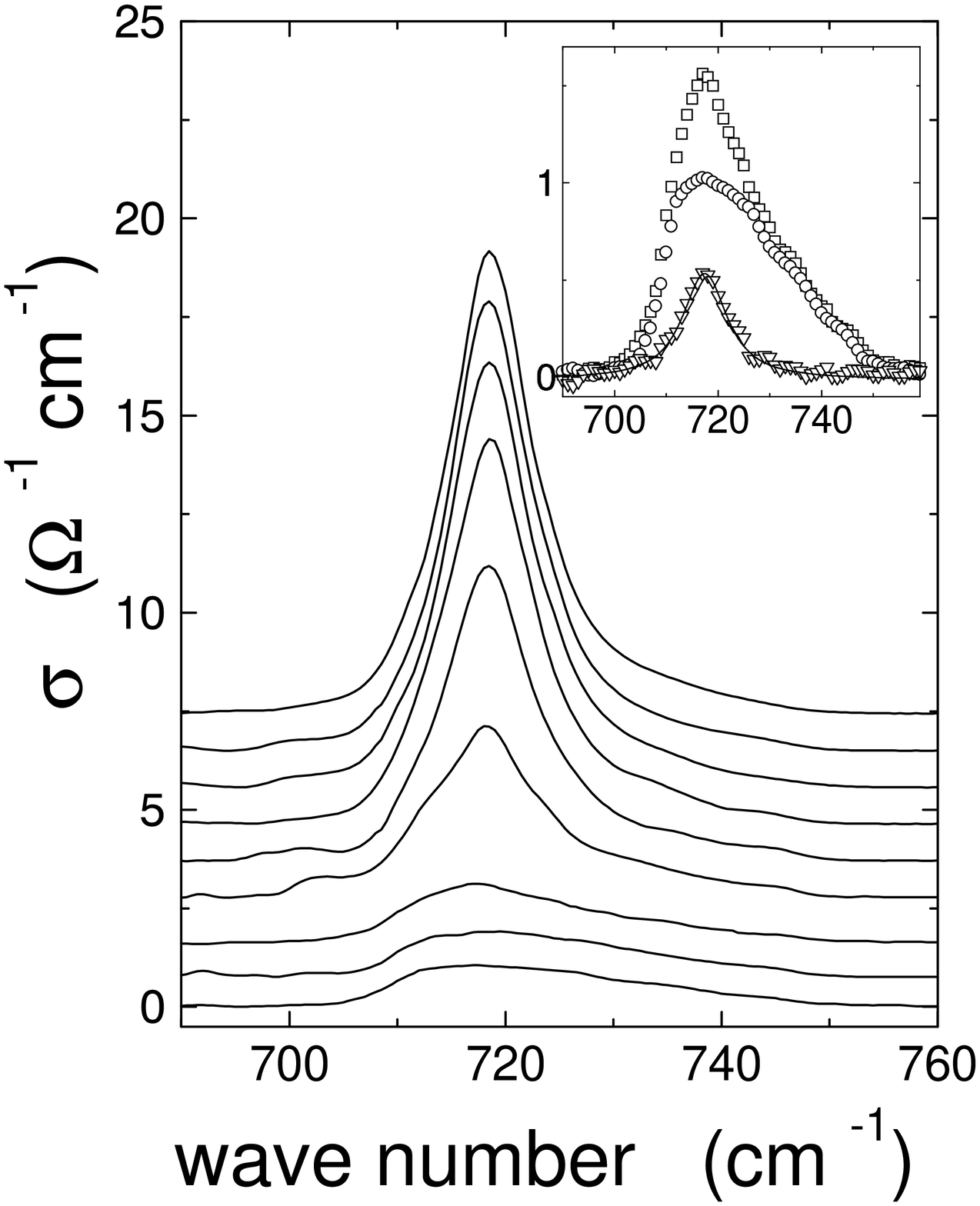,width=6cm,height=7cm}
\caption{Temperature evolution of the peak at 718~cm$^{-1}$ (temperatures from
up to down:  21.8, 24, 27.1, 29.5, 32, 33.5, 34.3, 35.3, 38 K) observed for
a ${\bf {b}}$-axis polarization. Spectra are vertically shifted
by 0.9 $\Omega^{-1}$cm$^{-1}$. The 
insert illustrates the correction procedure, as described
in the text ($\square $: 34.3K, $\bigcirc $: 38K, $\protect\nabla $:
difference, line: Lorentzian fit).}
\label{peak}
\end{center}
\end{figure}

\begin{figure}[tbp]
\begin{center}
\epsfig{file=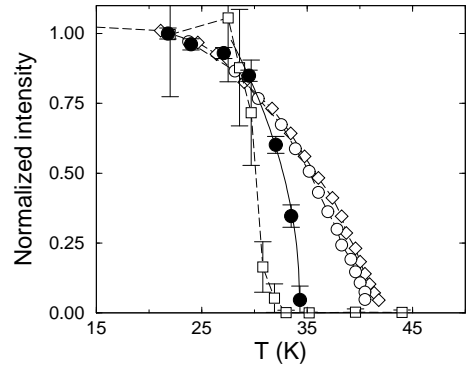,width=6cm}
\caption{Temperature dependence of the normalized squared dimerization
obtained by ED (14 ($\lozenge $), 16 ($\circ $) sites), QMC ($\square $, 40
sites) and intensity for the phonon mode observed at 718~cm$^{-1}$ in ${\bf {%
b}}$-axis polarization ($\bullet $). The solid line 
is a fit to a $(1-T/T_{SP})^{2%
\protect\beta }$ law. All data are normalized to their 22 K value.
Error bars are shown for the QMC and experimental
data.}
\label{intensity}
\end{center}
\end{figure}

We performed a careful investigation of the 718~cm$^{-1}$ line in the
conductivity over
temperatures between 20 and 38 K since this line is clearly associated to
the lattice transformation at $T<T_{SP}$ (Fig.~\ref{peak}). At low
temperatures (T=21.8, 24, 27 K) this peak is well described as a single
Lorentz line with a constant linewidth. On the other hand, above 29 K the
line profile is well described as superposition of a temperature-dependent
Lorentz peak on top of a temperature-independent asymmetric broad line which
can be clearly seen at 35.3 K and 38 K. This broad line might well be the
signature of structural fluctuations growing in the vicinity of $T_{SP}$. 
Indeed, after subtraction of the
temperature-independent part, a perfect Lorentz line with constant linewidth
(8.2$\pm $0.5)~cm$^{-1}$ is obtained for all T below the
transition. We believe that the Lorentz-type line corresponds to a folded
zone boundary phonon mode activated by the structural distortion and,
therefore, the intensity of this line should be proportional to the squared
dimerization. The
temperature dependence of the normalized line intensity is shown in Fig.~%
\ref{intensity} as well as the result of a fit with the 
$(1-T/T_{\mathrm{SP}})^{2\beta }$
law. This fit in the temperature
range 30--35 K gives $T_{\mathrm{SP}}$=(34.3$\pm $0.2) K, 
$\beta $=0.25$\pm $0.1. One should note that the relatively small error on
$\beta$ achieved with only few points near the $T_{\mathrm{SP}}$ is
due to the correction procedure described above together with the low
dispersion of experimental data as evident in Fig.~\ref{peak} (insert).

We proceed now to discuss the above results in the context of existing
theoretical approaches. The crudest description of the spin-Peierls ground
state (GS) is based on the dimerized Heisenberg model: the antiferromagnetic
coupling between nearest neighbor spins is assumed to be {\it statically}
modulated (see Eq. (\ref{h2}) below) along each spin-1/2 chain.
In fact, the physical origin of such a modulation relies on the 
{\it dynamical} magneto-elastic coupling.
One of the simplest model including dynamical effects of phonons read~\cite
{Khomskii1,affleck,augier3} (omitting the chain index), 
\begin{equation}
H=J\sum_{i}(1+\lambda u_{i}){\bf S}_{i}.{\bf S}_{i+1}+H_{{\rm ph}}^{0}+H_{%
{\rm elastic}}^{\perp }\ ,  \label{h1}
\end{equation}
where $\lambda $ is the magneto-elastic coupling constant and $u_{i}$
correspond to atomic displacements which modulate the
exchange integral $J$. Here, we restrict
ourselves to a single polarization of the phonon modes. Assuming
independent optical phonons, the phonon
hamiltonian takes the usual form, $H_{{\rm ph}}^{0}=\sum_{i}\{%
\frac{p_{i}^{2}}{2M}+\frac{1}{2}K_{\parallel }u_{i}^{2}\}$. An interchain
elastic coupling $H_{{\rm elastic}}^{\perp }$ is required to account for the
three-dimensional coherence of the phonons and involves an additional
elastic coupling $K_{\perp }$ between
neighboring chains. $H_{{\rm elastic}}^{\perp }$ is needed to obtain a finite
value of $T_{SP}$. Note that $K_{\parallel }$ and $%
\lambda $ depend strongly on the actual direction in space of the
displacements $u_{i}$ which is not explicitely specified in 
(\ref{h1}) (Ref.~\onlinecite{note}).

The optical conductivity is theoretically defined as the GS correlation $\sigma
(\omega)\propto\langle A_{IR}^\dagger(\omega+i0^+ -H)^{-1}A_{IR}\rangle_0$
where the operator $A_{IR}=\sum_i A_i$ describes the coupling between the ($%
{\bf q}=0$) photons and the medium. Quite generally, one expects $%
A_{i}\propto E^\alpha u_i^\beta T_{\alpha\beta}(i)$, where ${\bf E}$ is the
electric field of the light, ${\bf u}_i$ is the polarization of the phonon
mode and $T_{\alpha\beta}(i)$ is the IR tensor which might include local
spin fluctuations~\cite{lorenzana}.

In lowest order, $A_{IR}$ corresponds to the emission of a ${\bf q}=0$
phonon {\it i.e.} $T_{\alpha \beta }(i)=\delta _{\alpha \beta }$. According
to Ref.\onlinecite{Popova}, 
at $T>T_{\mathrm{SP}}$\ there are 15 ${\bf E\parallel a}$\ and
7 ${\bf E\parallel b}$\ phonons which should be active in the first order IR
processes. We did not observe all the
predicted modes probably because some of them have small oscillator
strengths. The group theoretical analysis in the SP phase is
not available yet.

\begin{figure}[hbt]
\begin{center}
\epsfig{file=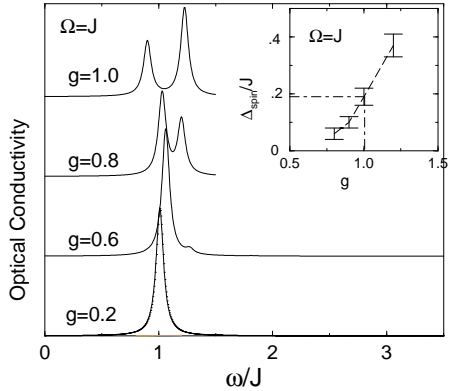,width=6cm}
\caption{ Optical conductivity vs frequency calculated within model (\ref{h1}%
) for $\Omega/J=1$ and for various spin-phonon coupling strengths $g$.
Insert: $\Delta_{\mathrm{spin}}$ vs g. }
\label{cond_theo}
\end{center}
\end{figure}

Calculated optical conductivity spectra ($T=0$) obtained by ED of
a N=12 site ring are shown
in Fig.~\ref{cond_theo} for a typical phonon frequency $\Omega =J$ and
several values of the dimensionless coupling constant 
$g=\lambda /\sqrt{2M\Omega }$ 
($H_{{\rm elastic}}^{\perp }$ is neglected here and a variational basis
is used for the phonons~\cite{augier3}). 
As shown in the insert, the spin gap (and hence the
dimerization) increases strongly with the coupling $g$. The doubling of the
unit cell leads to an additional peak in $\sigma (\omega )$ corresponding to
a new ${\bf q}=0$ phonon mode at higher energy resulting from a folding of
the Brillouin zone. This mechanism can then explain the experimental
observation of the new line at 718~cm$^{-1}$ if one assumes that the atomic
displacements $u_{i}$ have a component along 
the chains direction (${\bf b}$-axis).

As suggested theoretically~\cite{lorenzana} 
and observed experimentally~\cite
{Sr2CuO3}, higher order processes involving one phonon and two-magnon
scattering can also take place with, e.g.,
an IR tensor like $T_{\alpha
\beta }(i)={\bf S}_{i}\cdot {\bf S}_{i+1}(\delta _{\alpha \beta }
-{\bf e}_{\alpha }^{i,i+1}{\bf e}_{\beta }^{i,i+1})$, where 
${\bf e}^{i,i+1}$ is the
unit vector along the chain direction. In this case, the maximum
absorption would occur for $\bf E$ polarized along ${\bf
a}$. A continuous background is
indeed observed for such a geometry (see Fig.~\ref{refl}(a)) and could be
associated to a two magnon phonon 
assisted absorption. More detailed theoretical
and experimental investigations of this effect are left for a future study.

Although the treatment of the phonon dynamics at finite T is
beyond the scope of this work, a simpler mean field treatment of the lattice
is justified if the relevant phonons have a pronounced 3D character ({\it %
i.e.} $K_{\perp }$ is not too small compared to $K_{\parallel }$) and if 
$\Omega/J\lesssim1$. In the adiabatic approximation (which
corresponds, strictly speaking, to the limit $\Omega /J\rightarrow 0$) the
hamiltonian reads~\cite{riera}, 
\begin{equation}
H_{MF}=J\sum_{i}(1+(-1)^{i}\delta ){\bf S}_{i}\cdot {\bf S}_{i+1}+\frac{1}{2}%
NK\delta ^{2}\ ,  \label{h2}
\end{equation}
where $N$ is the number of sites, K an effective elastic constant and $%
\delta =\delta (T)$ the static dimerization which depends on the system
size. The value of $\delta (T)$ is obtained by minimizing the total free
energy $F_{T}=F_{{\em {spin}}}(T,\delta )+\frac{1}{2}NK\delta ^{2}$, where $%
F_{{\em {spin}}}$ corresponds to the spin part of hamiltonian (\ref{h2}). The
zero temperature value $\delta (0)\simeq 0.048$ is imposed by the actual
value of the ratio $\Delta _{{\em {spin}}}/J\simeq 0.2$ known from
experiments~\cite{Weiden}. The value of $\delta (T=0)$, in turn, enforces
the value of the ratio $K/J\simeq 3$. Note that $K$ and $J$ are the only
two free parameters of the model ($J\simeq440$~K is 
determined from a fit of the
high-T spin susceptibility\cite{Weiden}) and, hence, the theoretical 
value for $T_{\mathrm{SP}}$ comes as an output of the
model. The QMC calculation suggests $T_{\mathrm{SP}}\simeq33$~K, 
a value very
close to the experimental one. The explicit temperature dependence 
$\delta^2(T)$ obtained by
ED and QMC techniques\cite{note_ED} 
agrees reasonably with the experimental data, as shown in
Fig.~\ref{intensity}, apart from the value of the critical 
exponent~\cite{note_CuGeO}.

In summary, we have studied the infrared reflectivity on NaV$_{2}$O$_{5}$\
single crystals at temperatures down to 20K. The SP transition
order parameter was measured through the phonon spectrum temperature
dependence. These results are
compared with calculations based on mean field or dynamical treatments 
of the lattice.

We acknowledge A.~Damascelli and D. van der Marel for communicating their
results of a similar study prior to publication.


\begin{references}
\bibitem[{*}]{}  Permanent address: A.~F.~Ioffe Physical Technical
Institute, 194021 St. Petersburg, Russia

\bibitem{Hase}  M.~Hase {\it et al.}, Phys. Rev. Lett. {\bf 70}, 3651 (1993).

\bibitem{Isobe}  M.~Isobe and Y.~Ueda, J. Phys. Soc. Jpn. {\bf 65}, 1178
(1996).

\bibitem{Organic}  For a review on organic SP systems see e.g. J.~W.~Bray,
L.~V.~Interrante, I.~S.~Jacobs and J.~C.~Bonner, in {\it Extended Linear
Chain Compounds}, edited by J.~S.~Miller (Plenum Press, New York, 1983),
Vol. {\bf 3}.

\bibitem{RevCuGeO3}  For a review on CuGeO$_{3}$ see e.g. J.~P.~Boucher and
L.~P.~Regnault, J. Phys. I (Paris) {\bf 6}, 1939 (1996).

\bibitem{pouget} J.~P.~Pouget {\it et al.}, Phys. Rev. Lett. {\bf 72}, 4037 (1994).

\bibitem{regnault} L.~P.~Regnault {\it et al.}, Phys. Rev. B {\bf 53}, 5579 (1996).

\bibitem{X-rays}  Y.~Fujii {\it et al.}, J. Phys. Soc. Jpn. {\bf 66}, 326
(1997).

\bibitem{Light}  M.~Cottam and D.~Lockwood, {\it Light Scattering in
Magnetic Solids} (John Wiley \& Sons, New York, 1986)

\bibitem{OptCuGeO3}  P.~H.~M.~Van Loosdrecht, cond-mat/9711091 (unpublished).

\bibitem{li} G.~Li {\it et al.}, Phys. Rev. B {\bf 54}, R15 633 (1996).

\bibitem{Damasc}  A.~Damascelli {\it et al.}, Phys. Rev. B {\bf 56},
R11 373(1997).

\bibitem{Popova}  M.~N.~Popova {\it et al.}, Pis'ma v ZhETF {\bf 65}, 711
(1997).

\bibitem{Weiden}  M.~Weiden {\it et al.}, Z. Phys. B {\bf 103}, 1 (1997).

\bibitem{riera}  J.~Riera and A.~Dobry, Phys. Rev. B {\bf 51}, 16 098 (1995);
A.~Feiguin et al.,  Phys. Rev. B {\bf 56}, 14 607 (1997).

\bibitem{Castilla}  G.~Castilla {\it et al.}, Phys. Rev. Lett. {\bf 75},
1823 (1995).

\bibitem{augier1}  D. Augier, D. Poilblanc, S. Haas, A. Delia and E.
Dagotto, Phys. Rev. B {\bf 56}, R5732 (1997).

\bibitem{Techno}  M.~Isobe {\it et al.}, submitted to J. Cryst. Growth
(1998).

\bibitem{Khomskii1}  D.~Khomskii, W.~Geerstma and M.~Mostovoy, Czech. J. of
Phys. {\bf 46}, Suppl S6, 32 (1996).

\bibitem{affleck}  I.~Affleck, proceedings of the NATO ASI ``{\it Dynamical
Properties of Unconventional Magnetic Systems}'', April 97, preprint
cond-mat/9705127, to be published.

\bibitem{augier3}  D.~Augier, D.~Poilblanc, E.~S\o rensen and I.~Affleck,
submitted to Phys. Rev. Lett. For results using a truncation procedure in
momentum space see also D.~Augier and D.~Poilblanc, Eur. Phys. J. B {\bf 1},
19 (1998).

\bibitem{note}  The dimerization $\delta$ can be defined within this
model by $\delta^2=\lambda^2\lim_{N\rightarrow\infty} \frac{1}{N}\sum_r
\langle u_i u_{i+r} \rangle (-1)^r$.

\bibitem{lorenzana}  T. Moriya, J. Appl. Phys. {\bf 39}, 1042 (1968),
J.~Lorenzana and G.~A.~Sawatzky, Phys. Rev. B {\bf 52},
9576 (1995) and J.~Lorenzana and R.~Eder, Phys. Rev. B {\bf 55}, R3358 (1997).

\bibitem{Sr2CuO3}  H.~Suzuura, H.~Yasuhara, A.~Furusaki, N.~Nagaosa and
Y.~Tokura, Phys. Rev. Lett. {\bf 76}, 2579 (1996).

\bibitem{note_ED} ED deals with smaller rings than QMC and, hence, has
larger finite size effects and therefore overestimate $T_{\mathrm{SP}}$.

\bibitem{note_CuGeO} Deviations from the mean-field behavior ($\beta=0.5$)
have also been seen in CuGeO$_3$; see e.g. Ref. \onlinecite{regnault}, 
\onlinecite{Damasc} 
and M.C.~Martin et al., Phys. Rev. B {\bf 53}, 14713 (1996). 


\end{references}
\end{document}